\documentclass[12pt]{article}
\usepackage{amsmath,amsfonts,amssymb,graphicx}
\usepackage{graphics, setspace}
\usepackage{cite} 
\usepackage{hyperref}
\usepackage{epstopdf}
\DeclareGraphicsExtensions{.eps}

\begin{document}
\thispagestyle{empty}

\def\theequation{\arabic{section}.\arabic{equation}}
\def\a{\alpha}
\def\b{\beta}
\def\g{\gamma}
\def\d{\delta}
\def\dd{\rm d}
\def\e{\epsilon}
\def\ve{\varepsilon}
\def\z{\zeta}
\def\B{\mbox{\bf B}}\def\cp{\mathbb {CP}^3}

\newcommand{\h}{\hspace{0.5cm}}

\begin{titlepage}

\renewcommand{\thefootnote}{\fnsymbol{footnote}}
\begin{center}
{\Large \bf Currents algebra for the two-sites Bose-Hubbard model}
\end{center}
\vskip 1.2cm \centerline{\bf Gilberto N. Santos Filho$^{1}$}

\vskip 10mm
\centerline{\sl$\ ^1$ Centro Brasileiro de Pesquisas F\'{\i}sicas - CBPF}
\centerline{\sl Rua Dr. Xavier Sigaud, 150, Urca, Rio de Janeiro - RJ - Brazil}
\vskip .5cm

\centerline{\tt gfilho@cbpf.br}
 
\vskip 20mm

\baselineskip 18pt

\begin{center}
{\bf Abstract}
\end{center}

I present a currents algebra for the two-sites Bose-Hubbard model, generalize the Heisenberg equation of motion to write the second time derivative of the currents operators and use it to get the quantum dynamics of the currents. For different choices of the Hamiltonian parameters I get different currents dynamics and determine the period of the oscillations in function of the parameters.

\end{titlepage}
\newpage
\baselineskip 18pt

\def\nn{\nonumber}
\def\tr{{\rm tr}\,}
\def\p{\partial}
\newcommand{\non}{\nonumber}
\newcommand{\bea}{\begin{eqnarray}}
\newcommand{\eea}{\end{eqnarray}}
\newcommand{\bde}{{\bf e}}
\renewcommand{\thefootnote}{\fnsymbol{footnote}}
\newcommand{\be}{\begin{eqnarray}}
\newcommand{\ee}{\end{eqnarray}}

\vskip 0cm

\renewcommand{\thefootnote}{\arabic{footnote}}
\setcounter{footnote}{0}

\setcounter{equation}{0}
\section{Introduction}

 The early experimental realization of a two-wells Bose-Einstein condensate (BEC) was made only two years after the experimental verification of the BEC  \cite{ak,anderson,wwcch,bose,eins,Hulet,Dalfovo,l01,Bagnato,cw,Donley,Piza,Bloch,Carusotto} to study the interference between two freely expanding condensates \cite{Ketterle1,Ketterle2}, and their results had direct implications in the study of the atom laser and the Josephson effect \cite{Josephson1,Josephson2} for BEC. Some models were used to study some behaviors of these systems as for example the quantum phase transitions,  the classical analysis and the quantum dynamics \cite{GSantos06a,GSantos09,GSantos10,sarma,Vely}. I am considering here  the two-sites Bose-Hubbard model, also known as the {\it Canonical Josephson Hamiltonian} \cite{l01}, that has been an useful model in understanding tunneling phenomena using two BECs   \cite{Javanainen,milb,l02,hines2,ours,our,hines}.  This model is integrable in the sense that it can be solved by the quantum inverse scattering method (QISM) \cite{GSantos11a,GSantos11b,jlletter,jlreview,jlsigma,Angela,ATonel,GSantos06b,Angela2,GSantosBVS-arxiv,jdensity} and it has been discussed in different ways using this method  \cite{jlletter,jlreview,jlsigma,Angela,ATonel,GSantos06b,Angela2,GSantosBVS-arxiv}.  In this context this model is a particular case of the bosonic multi-state  model studied in \cite{GSantos13}.  The experimental quantum dynamics and the classical analysis of this model was performed by \cite{Oberthaler1,Oberthaler2,Oberthaler3}.  In this letter I will discuss the currents algebra for the two-sites Bose-Hubbard  model.  Currents algebra was introduced by M. Gell-Mann in high energy physics to study partially conserved axial vector current in the beta decay \cite{Gell}. 
I generalize  the Heisenberg equation of motion to write the second time derivative of any operator and use it to study the quantum dynamics of the currents. This method can be applied to many systems that present microscopic tunneling phenomenon to get some characteristic energies of the systems in function of the period of the oscillation and  that is also important to quantum metrology \cite{QMet1,QMet2,QMet3,QMet4,QMet5,QMet6,QMet7}. The model is described by the Hamiltonian 
\begin{equation}
\hat{H} = \frac{K}{8}(\hat{N}_1 - \hat{N}_2)^2 - \frac{\Delta \mu}{2}(\hat{N}_1 -\hat{N}_2)
 - \frac{\mathcal{E}_J}{2}(\hat{a}_1^\dagger \hat{a}_2 + \hat{a}_2^\dagger \hat{a}_1), 
\label{ham} 
\end{equation}
\noindent  where $\hat{a}_1^\dagger$, $\hat{a}_2^\dagger$, denote the single-particle creation
 boson operators in the two wells and  $\hat{N}_1 = \hat{a}_1^\dagger \hat{a}_1,
 \hat{N}_2 = \hat{a}_2^\dagger \hat{a}_2$, are the corresponding number of particles 
 boson operators. These bosons operators satisfies the canonical commutation relations $[\hat{a}_i,\hat{a}_j^\dagger] = \delta_{ij}\hat{I}$, $[\hat{a}_i,\hat{a}_j]=[\hat{a}_i^\dagger,\hat{a}_j^\dagger] = 0$  and 
 $[\hat{N}_i,\hat{a}_j] =  - \delta_{ij}\hat{a}_j$, $[\hat{N}_i,\hat{a}_j^\dagger] =  +\delta_{ij}\hat{a}_j^\dagger$, where $\hat{I}$ is the identity operator. The coupling $K$ provides the strength of the $s$-wave scattering interaction between the bosons,  $\Delta \mu$ is the external potential and ${\cal E}_J$ is the amplitude of tunneling.

\section{Symmetries} 

 The Hamiltonian (\ref{ham}) is  invariant under the $\mathbb{Z}_2$ mirror transformation $\hat{a}_j \rightarrow -\hat{a}_j, \hat{a}_j^{\dagger} \rightarrow -\hat{a}_j^{\dagger}$, and under the global $U(1)$  gauge transformation $\hat{a}_j \rightarrow e^{i\alpha}\hat{a}_j$, where $\alpha$ is an arbitrary $c$-number and $\hat{a}^{\dagger}_j\rightarrow e^{-i\alpha}\hat{a}^{\dagger}_j,\;\;j=1,2$. For $\alpha = \pi$ we get again the $\mathbb{Z}_2$ symmetry. The global $U(1)$  gauge invariance is associated with the conservation of the total number of atoms $\hat{N} = \hat{N}_1 + \hat{N}_2$ and the $\mathbb{Z}_2$ symmetry is associated with the parity of the wave function by the relation $\hat{P} \; |\Psi\rangle = (-1)^N |\Psi\rangle$, with  

\begin{equation}
|\Psi\rangle = \sum_{n = 0}^N \; C_{n,N-n} \frac{(\hat{a}_1^{\dagger})^{n}}{\sqrt{n!}} \frac{(\hat{a}_2^{\dagger})^{N - n}}{\sqrt{(N - n)!}} |0,0\rangle, \label{wf1}
\end{equation} 
\noindent where $\hat{P}$ is the parity operator and $[\hat{H},\hat{P}]=0$.

\begin{figure}
\begin{center}
\includegraphics[scale=0.25]{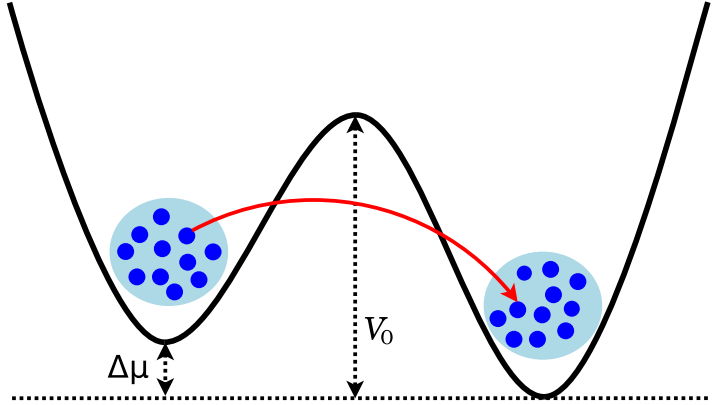}
\caption{(Color online) A two-wells potential representation of the two-sites Bose-Hubbard model for the case   $\Delta\mu \neq 0$ and barrier height $V_0$. We are considering one condensate, in light blue, for each well. We show one atom, in dark blue, tunneling from left to right (red arrow). }
\label{tw1}
\end{center}
\end{figure}

 There is also the permutation symmetry of the atoms of the two wells if we have $\Delta\mu = 0$ and when we turn on $\Delta\mu$ we break the symmetry. The wave function (\ref{wf1}) is symmetric under this permutation
 \begin{equation}
 \hat{{\cal P}}\; |\Psi\rangle = \sum_{n = 0}^N \;  C_{N - n,n} \frac{(\hat{a}_1^{\dagger})^{N - n}}{\sqrt{(N - n)!}}\frac{(\hat{a}_2^{\dagger})^{n}}{\sqrt{n!}} |0,0\rangle = |\Psi\rangle,
\end{equation}  
\noindent where $\hat{{\cal P}}$ is the permutation operator and $[\hat{H},\hat{{\cal P}}]=0$ if $\Delta\mu = 0$ \cite{jlletter}.  In the antisymmetric case $\Delta\mu\neq0$ we can change the bias of one well. In this case it is called a tilted two-wells potential \cite{ours,Dounas}.  In the Fig. (\ref{tw1}) we represent the two BECs by a two-wells potential for the case  $\Delta\mu \neq 0$. We  get the two-sites Bose-Hubbard model when we consider each  BEC as a site. 

\section{Currents Algebra}

The total  particles number boson operator, $\hat{N} = \hat{N}_1+\hat{N}_2$, 
is a conserved quantity and it is a commutable compatible operator  with the particles number bosons operators in each well, $[\hat{N},\hat{N}_1]=[\hat{N},\hat{N}_2]= [\hat{N}_1,\hat{N}_2]=0$. The number of particles bosons operators  in each well don't commute with the Hamiltonian and their time evolution is dictated by the Josephson tunneling current operator, $\hat{\mathcal{J}} = \frac{1}{2i} (\hat{a}_1^\dagger \hat{a}_2 - \hat{a}_2^\dagger \hat{a}_1)$,  in coherent opposite phases because of the conservancy of $\hat{N}$. We get the following equations for the time evolution

\begin{eqnarray}
\hat{N}_1(t)  &=&  \hat{N}_1(0) - \frac{\mathcal{E}_J}{\hbar}\int_0^t\hat{\mathcal{J}}(\tau) \; d\tau, \label{SCDN1} \\
\hat{N}_2(t) &=& \hat{N}_2(0) + \frac{\mathcal{E}_J}{\hbar}\int_0^t\hat{\mathcal{J}}(\tau) \; d\tau. \label{SCDN2}
\end{eqnarray}

Here is worth to note that the two BECs are entangled by the tunneling of the particles and we can study the quantum phase transition of the system using tools of the quantum information \cite{GSantos09,GSantos10}.

   The tunneling current $\hat{\mathcal{J}}$ together with the imbalance current $\hat{\mathcal{I}} = \frac{1}{2}(\hat{N}_1 - \hat{N}_2)$ and the coherent correlation tunneling current operator $\hat{\mathcal{T}} = \frac{1}{2}(\hat{a}_1^\dagger \hat{a}_2 + \hat{a}_2^\dagger \hat{a}_1)$, generate together the currents algebra $[\hat{\mathcal{T}},\hat{\mathcal{J}}]= +i\hat{\mathcal{I}}$, $[\hat{\mathcal{T}},\hat{\mathcal{I}}]= -i\hat{\mathcal{J}}$ and $[\hat{\mathcal{J}},\hat{\mathcal{I}}]= +i\hat{\mathcal{T}}$. With the identification $\hat{L}_x \equiv  \hbar\hat{\mathcal{T}}$, $\hat{L}_y \equiv  \hbar\hat{\mathcal{J}}$, and $\hat{L}_z \equiv \hbar\hat{\mathcal{I}}$ we can write this currents algebra in the standard  compact way of the momentum angular algebra  $[\hat{L}_k,\hat{L}_l] = i \hbar \varepsilon_{klm}\hat{L}_m$, where $\varepsilon_{klm}$ is the  antisymmetric Levi-Civita tensor with $k,l,m=x,y,z$ and $\varepsilon_{xyz} = +1$.  We have two Casimir operators for that currents algebra. One of them is the total number of particles, $\hat{\mathcal{\mathfrak{C}}}_{1} = \hat{N}$, related to the $U(1)$ symmetry and the another one is related to the momentum angular algebra and the $O(3)$ symmetry, $\hat{\mathcal{\mathfrak{C}}}_{2} = \hat{\mathcal{T}}^2 + \hat{\mathcal{I}}^2 + \hat{\mathcal{J}}^2$. It is direct to show that $\hat{\mathcal{\mathfrak{C}}}_{2}$ is just a function of $\hat{\mathcal{\mathfrak{C}}}_{1}$ 
 
\begin{equation}
\hat{\mathcal{\mathfrak{C}}}_{2} = \frac{\hat{\mathcal{\mathfrak{C}}}_{1}}{2}\left(\frac{\hat{\mathcal{\mathfrak{C}}}_{1}}{2} +  1\right).
\end{equation}
\noindent and that the Casimir surface is spherical with radius $\sqrt{\langle\hat{\mathcal{\mathfrak{C}}}_{2}\rangle}$.

\section{Currents Quantum Dynamics}

We can rewrite the Hamiltonian (\ref{ham}) using the currents operators
\begin{equation}
\hat{H} = \frac{K}{2}\hat{\mathcal{I}}^2 - \Delta\mu\hat{\mathcal{I}} 
 - \mathcal{E}_J\hat{\mathcal{T}}.
\label{ham2} 
\end{equation}

The quantum dynamic of the currents are determined by the currents algebra, their commutation relations with the Hamiltonian and the parameters. If the Hamiltonian is not explicitly time-dependent it is not time-dependent, $\frac{d\hat{H}}{dt} =  \frac{\partial\hat{H}}{\partial t} = 0 $, and the system is closed (conservative). It is also important to note that the Hamiltonian is the same in the Heisenberg and Schr\"odinger pictures, $\hat{H}_H = \hat{H}_S$. Using these facts we can  write the second time derivative of any operator $\hat{O}$ in the Heisenberg picture as
\begin{equation}
 \frac{d^2\hat{O}}{dt^2} = \left(\frac{i}{\hbar}\right)^2 [\hat{H},[\hat{H},\hat{O}]],
 \label{dOdt0}
\end{equation}
\noindent or as
\begin{equation}
 \frac{d^2\hat{O}}{dt^2} = \frac{i}{\hbar}[\hat{H},\frac{d\hat{O}}{dt}].
 \label{dOdt}
\end{equation}

It is direct to generalize the Eqs. (\ref{dOdt0}) and (\ref{dOdt}) for higher time derivatives. We can get similar equation in the another pictures. The pictures preserve the commutation relations between the operators in the sense that if we have $[\hat{A}_S,\hat{B}_S]=\hat{C}_S$ in the Schr\"odinger picture we get the same relation in the Heisenberg picture, $[\hat{A}_H,\hat{B}_H]=\hat{C}_H$, and in the interaction picture, $[\hat{A}_I,\hat{B}_I]=\hat{C}_I$. The same is true for the anticommutators, and so the pictures preserve the algebra. We can see from Eqs. (\ref{ham2}) and (\ref{dOdt0}) that the Casimir operators $\hat{\mathcal{\mathfrak{C}}}_{1}$ and  $\hat{\mathcal{\mathfrak{C}}}_{2}$ are also  conserved quantities, $[\hat{H},\hat{\mathcal{\mathfrak{C}}}_{1}]=[\hat{H},\hat{\mathcal{\mathfrak{C}}}_{2}]=0$. Using the Eq. (\ref{dOdt0}) or (\ref{dOdt}) we found the following equations for the quantum dynamics of the three currents

\begin{equation}
\frac{d^{2}\hat{\mathcal{I}}}{dt^{2}} + \frac{\mathcal{E}_J^{2}}{\hbar^2}\hat{\mathcal{I}}  =  -\frac{\mathcal{E}_JK}{\hbar^{2}}\hat{\mathcal{I}}\hat{\mathcal{T}} + \frac{\mathcal{E}_J\Delta\mu}{\hbar^{2}}\hat{\mathcal{T}} + i\frac{\mathcal{E}_J K}{2\hbar^{2}}\hat{\mathcal{J}}, 
\label{eq1:wideeq}  
\end{equation}

\begin{eqnarray}
\frac{d^{2}\hat{\mathcal{J}}}{dt^{2}}+\frac{1}{\hbar^{2}}\left[(\Delta\mu)^{2} + \mathcal{E}_J^{2} +\frac{K^{2}}{4}\right]\hat{\mathcal{J}} &=&  -\frac{K^{2}}{\hbar^{2}}\hat{\mathcal{I}}^{2}\hat{\mathcal{J}} 
 - i\frac{K^{2}}{\hbar^{2}}\hat{\mathcal{I}}\hat{\mathcal{T}}  
 +  2\frac{K\Delta\mu}{\hbar^{2}}\hat{\mathcal{I}}\hat{\mathcal{J}} \nonumber \\ &-& \frac{K\mathcal{E}_J}{\hbar^{2}}\hat{\mathcal{J}}\hat{\mathcal{T}} +  i\frac{K\Delta\mu}{\hbar^{2}}\hat{\mathcal{T}} - \frac{i}{2}\frac{K\mathcal{E}_J}{\hbar^{2}}\hat{\mathcal{I}}, 
\label{eq2:wideeq}
\end{eqnarray}

\begin{eqnarray}
\frac{d^{2}\hat{\mathcal{T}}}{dt^{2}} + \frac{1}{\hbar^{2}}\left[(\Delta\mu)^{2} + \frac{K^{2}}{4}\right]\hat{\mathcal{T}}  &=&  -\frac{K^{2}}{\hbar^{2}}\hat{\mathcal{I}}^2\hat{\mathcal{T}} + i\frac{K^{2}}{\hbar^{2}}\hat{\mathcal{I}}\hat{\mathcal{J}} 
+ 2\frac{K\Delta\mu}{\hbar^{2}}\hat{\mathcal{I}}\hat{\mathcal{T}} \nonumber \\ 
&-&  \frac{K\mathcal{E}_J}{\hbar^{2}}\hat{\mathcal{I}}^{2} + \frac{K\mathcal{E}_J}{\hbar^{2}}\hat{\mathcal{J}}^{2} + \frac{\Delta\mu\mathcal{E}_J}{\hbar^{2}}\hat{\mathcal{I}} - i\frac{K\Delta\mu}{\hbar^{2}}\hat{\mathcal{J}}.
\label{eq3:wideeq}
\end{eqnarray}

\begin{figure}[t]
\begin{center}
\includegraphics[scale=0.7]{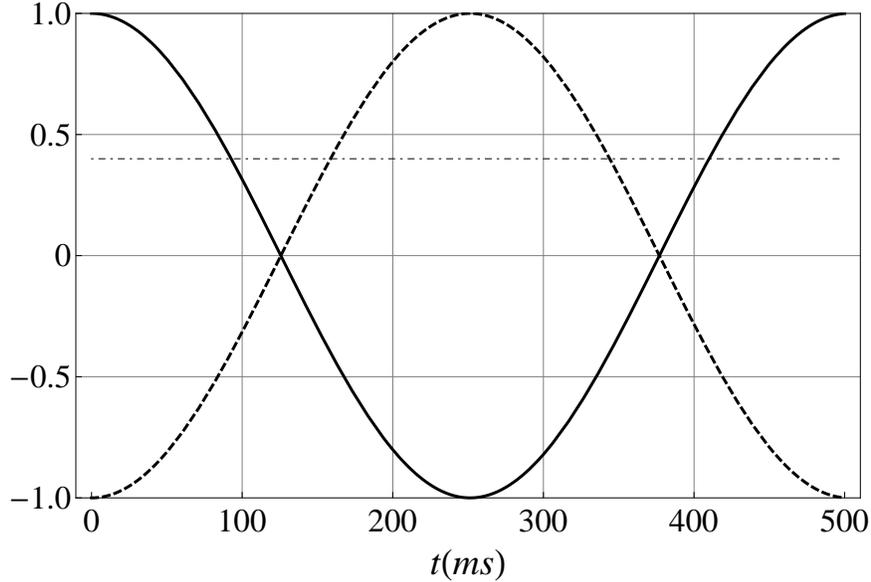} 
\caption{Quantum dynamics  of the currents for $\omega_{\mathcal{I}} = \omega_{\mathcal{J}} = 12.5\;$rad$\cdot$Hz. The initial condition for the current $\hat{\mathcal{I}}(t)$ (full line) is $\hat{\mathcal{I}}(0) = 1.0$. The initial condition for the current $\hat{\mathcal{J}}(t)$ (dashed line) is $\hat{\mathcal{J}}(0) = -1.0$. The initial condition for the current $\hat{\mathcal{T}}(t)$ (dot-dashed line) is $\hat{\mathcal{T}}(0) = 0.4$. The initial conditions for the first derivative of all currents is zero.}
\label{tw2}
\end{center}
\end{figure}

 We can see from  Eqs. (\ref{eq1:wideeq}), (\ref{eq2:wideeq}) and (\ref{eq3:wideeq}) that the currents are coupled on the right hand side of these equations.  Different choices of the ratio between the parameters of the Hamiltonian gives us different dynamics for the currents. In the Rabi regime, $K/\mathcal{E}_J\ll N^{-2}$ \cite{l01,l02,GSantos06b}.  Consequently, in the extreme Rabi regime we can neglect $K$ and consider the no interaction limit $K\rightarrow 0$.  Considering the symmetric case, $\Delta\mu = 0$, the current $\hat{\mathcal{T}}$ is a conserved  quantity, $[\hat{\mathcal{H}},\hat{\mathcal{T}}]=0$, but this don't means that we don't have tunneling. We can see from Eqs. (\ref{SCDN1}) and (\ref{SCDN2}) that the quantum dynamic of $\hat{N}_1$, $\hat{N}_2$, and $\hat{\mathcal{I}}$ only depend of the current $\hat{\mathcal{J}}$ and the amplitude of tunneling $\mathcal{E}_J$. Here is worth to note that in $\mathcal{E}_J$ is included the kinetic energy of the atoms. The current dynamics for these currents are the dynamic of the simple harmonic oscillator (SHO)

\begin{eqnarray}
\frac{d^2\hat{\mathcal{I}}}{dt^2} + \omega_{\mathcal{I}}^2\hat{\mathcal{I}} & =  & 0, \label{CD3} \\
\frac{d^2\hat{\mathcal{J}}}{dt^2} + \omega_{\mathcal{J}}^2\hat{\mathcal{J}} & =  & 0, \label{CD4}
\end{eqnarray}
\noindent where $\omega_{\mathcal{I}} = \omega_{\mathcal{J}} = \omega_{\mathcal{E}_J} = \frac{\mathcal{E}_J}{\hbar}$ is the natural angular frequency of the SHO. The period of the oscillations is 
$T = \frac{2\pi}{\omega_{\mathcal{E}_J}}$ and we get the following relation $\mathcal{E}_J T = h$ between the energy of the amplitude of tunneling and the period.   In the no interaction limit is expected a period of $T = 500\;$ms instead the period of $T = 40.1\;$ms for the interacting nonlinear regime as in the experiment \cite{Oberthaler1} and in the generalized model \cite{GSantosCA2}. We have gotten the value  $\mathcal{E}_J =  12.5\hbar\;$J for the amplitude of tunneling. In the Fig. (\ref{tw2}) we show the  solutions for the Eqs. (\ref{CD3}) and (\ref{CD4}). The currents are uncorrelated now  and there is no interference between them. We have Rabi dynamics for the currents $\hat{\mathcal{I}}$ and $\hat{\mathcal{J}}$ and  self-trapping  for the current $\hat{\mathcal{T}}$.

\begin{figure}[t]
\begin{center}
\includegraphics[scale=0.7]{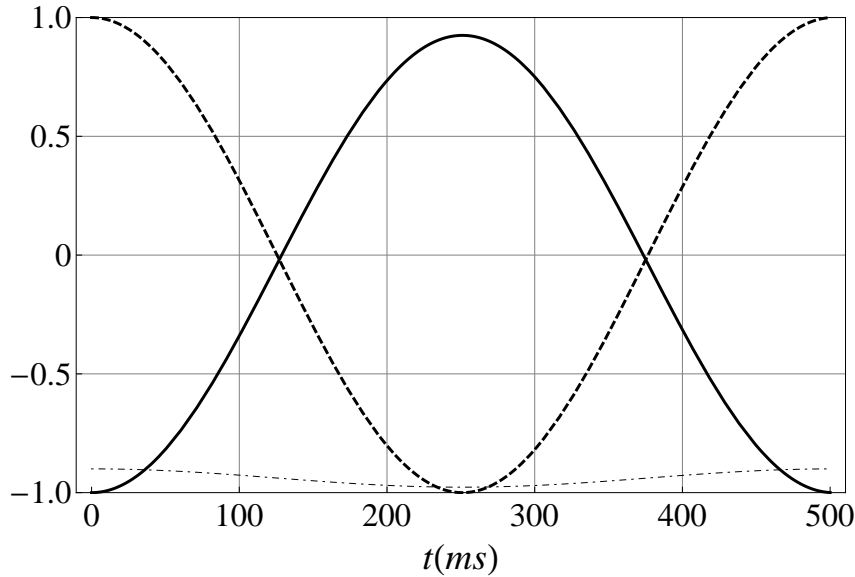} 
\caption{Quantum dynamics of the currents for $\omega_{\mathcal{E}_J} = 12.5\;$rad$\cdot$Hz and $\omega_{\Delta\mu} = 0.5\;$rad$\cdot$Hz. The initial condition for the current $\hat{\mathcal{I}}(t)$ (full line) is $\hat{\mathcal{I}}(0) = -1.0$. The initial condition for the current $\hat{\mathcal{J}}(t)$ (dashed line) is $\hat{\mathcal{J}}(0) = 1.0$. The initial condition for the current $\hat{\mathcal{T}}(t)$ (dot-dashed line) is $\hat{\mathcal{T}}(0) = -0.9$. The initial conditions for the first derivative of all currents is zero.}
\label{tw3}
\end{center}
\end{figure}

Breaking the symmetry, $\Delta\mu \neq 0$, to consider the antisymmetric case the currents dynamics are

\begin{equation}
\frac{d^2\hat{\mathcal{I}}}{dt^2} + \left(\frac{\mathcal{E}_J}{\hbar}\right)^2 \hat{\mathcal{I}} = \frac{\mathcal{E}_J\Delta\mu}{\hbar^2}\hat{\mathcal{T}},
\label{DDCD1}
\end{equation}

\begin{equation}
\frac{d^2\hat{\mathcal{T}}}{dt^2} + \left(\frac{\Delta\mu}{\hbar}\right)^2 \hat{\mathcal{T}} = \frac{\mathcal{E}_J\Delta\mu}{\hbar^2}\hat{\mathcal{I}},
\label{DDCD2}
\end{equation}

\begin{equation}
\frac{d^2\hat{\mathcal{J}}}{dt^2} + \frac{(\Delta\mu)^2 + \mathcal{E}_J^2}{\hbar^2}\hat{\mathcal{J}} = 0.
\label{DDCD3}
\end{equation}
\noindent The Eq. (\ref{DDCD3}) describes a SHO with natural angular frequency $\omega_{\mathcal{J}} = \sqrt{\omega_{\Delta\mu}^2 + \omega_{\mathcal{E}_J}^2}$,  period of the oscillations $T = \frac{2\pi}{\sqrt{\omega_{\Delta\mu}^2 + \omega_{\mathcal{E}_J}^2}}$ and  relation $ET = h$, with $E = \hbar\omega_{\mathcal{J}}$, between the period and the energy. The Eqs. (\ref{DDCD1}) and (\ref{DDCD2}) are a system of two second order linear differential equations. If we diagonalize the matrix of the coefficients we get the same frequency $\omega_{\mathcal{J}}$.   In the Fig. (\ref{tw3}) we show the numeric solution for the Eqs. (\ref{DDCD1}), (\ref{DDCD2}) and (\ref{DDCD3}). We choose the same period, $T = 500\;$ms to get the values  $\mathcal{E}_J =  12.5\hbar\;$J for the amplitude of tunneling and $\Delta\mu = 0.5\hbar\;$J for the external potential. We have Rabi dynamics for the currents $\hat{\mathcal{J}}$, Josephson dynamics for the current $\hat{\mathcal{I}}$ and self-trapping  for the current $\hat{\mathcal{T}}$.  The currents $\hat{\mathcal{I}}$ and $\hat{\mathcal{T}}$ are correlated and there is interference between them.

\section{Summary}
In summary, I have showed that a currents algebra appears when we calculate the quantum dynamics of the number bosons operators of each well. I have generalized the Heisenberg equation of motion to write the second time  derivative of any operator. Then I have calculated the quantum dynamics of these currents and have showed that different dynamics appear when we consider different choices of the parameters of the Hamiltonian.  For specific choices of the parameters some of the currents are uncorrelated and there is no interference between them. 

\section*{Acknowledgement}
I acknowledge CAPES/FAPERJ (Coordena\c{c}\~ao de Aperfei\c{c}oamento de Pessoal de N\'{\i}vel Superior/Funda\c{c}\~ao de Amparo \`a Pesquisa do Estado do Rio de Janeiro) for the financial support.


\end{document}